\documentclass[fleqn,usenatbib]{mnras}

\usepackage{newtxtext,newtxmath}
\usepackage{booktabs, tabularx}

\usepackage[T1]{fontenc}

\DeclareRobustCommand{\VAN}[3]{#2}
\let\VANthebibliography\thebibliography
\def\thebibliography{\DeclareRobustCommand{\VAN}[3]{##3}\VANthebibliography}

\usepackage{graphicx}	
\usepackage{amsmath}	
\newcommand{\fermi}{{\it Fermi}-LAT}
\newcommand{\gray}{$\gamma$-ray}
\newcommand{\grays}{$\gamma$-rays}
\newcommand{\source}{PKS 0537-286}

\title[Broadband emission from PKS 0537-286]{Origin of multiwavelength emission from flaring high redshift blazar PKS 0537-286}

\author[N. sahakyan et al.]{
N. Sahakyan$^{1,2, 3}$, \thanks{E-mail: narek@icra.it}
G. Harutyunyan$^{1, 4}$,
D. Israyelyan$^{1}$\\
$^{1}$ICRANet-Armenia, Marshall Baghramian Avenue 24a, Yerevan 0019, Armenia\\
$^{2}$ICRANet, P.zza della Repubblica 10, 65122 Pescara, Italy\\
$^{3}$ ICRA, Dipartimento di Fisica, Sapienza Universita` di Roma, P.le Aldo Moro 5, 00185 Rome, Italy\\
$^{4}$ Byurakan Astrophysical Observatory, Aragatsotn reg., Armenia
}

\date{Accepted XXX. Received YYY; in original form ZZZ}

\pubyear{2015}

\begin{document}
\label{firstpage}
\pagerange{\pageref{firstpage}--\pageref{lastpage}}
\maketitle

\begin{abstract}
The high redhsift blazars powered by supermassive black holes with masses exceeding $10^9\:M_\odot$ have the highest jet power and luminosity and are important probes to test the physics of relativistic jets at the early epochs of the Universe. We present a multi-frequency spectral and temporal study of high redshift blazar PKS 0537-286 by analyzing data from \fermi, NuSTAR Swift XRT and UVOT. Although the time averaged \gray\ spectrum of the source is relatively soft (indicating the high-energy emission peak is below the GeV range), several prominent flares were observed when the spectrum hardened and the luminosity increased above $10^{49}\:{\rm erg\:s^{-1}}$. The X-ray emission of the source varies in different observations and is characterised by a hard spectrum $\leq1.38$ with a luminosity of $>10^{47}\:{\rm erg\:s^{-1}}$. The broadband spectral energy distribution in the quiescent and flaring periods was modeled within a one-zone leptonic scenario assuming different locations of the emission region and considering both internal (synchrotron radiation) and external (from the disk, broad-line region and dusty torus) photon fields for the inverse Compton scattering. The modeling shows that the most optimistic scenario, from the energy requirement point of view, is when the jet energy dissipation occurs within the broad-line region. The comparison of the model parameters obtained for the quiescent and flaring periods suggests that the flaring activities are most likely caused by the hardening of the emitting electron spectral index and shifting of the cut-off energy to higher values.
\end{abstract}




\section{Introduction}
Blazars are radio-loud quasars with powerful relativistic jets that make a small angle to the observer's line of sight \citep{1995PASP..107..803U}. They are among the most energetic sources in the Universe and a dominant class of extragalactic sources in the high energy (HE; $>100$ MeV) \gray\ sky \citep[e.g.,][]{2022ApJS..260...53A}. The new possibility of extensive multiwavelength observations coupled with multi-messenger observations have the potential to widen our understanding of blazars.

Historically, blazars are sub-grouped in two large sub-classes: BL Lacs and flat-spectrum radio quasars (FSRQs) \citep{1995PASP..107..803U}. BL Lacs have nearly featureless optical spectra (very weak or no lines are observed) while the FSRQs have bright and broad lines with equivalent widths of ${\rm |EM|>}$ 5\:{\AA}. One of the most distinguishable features of blazars is the very strong variability of their emission in almost all the observed bands in various times scales, from minute to week or month scales; \citep[e.g.,][]{2013ApJ...762...92A, 2014Sci...346.1080A, 2016ApJ...824L..20A, 2018ApJ...854L..26S, 2019ApJ...877...39M}. This variability is stochastic in nature but a recent detection of quasi-periodic oscillations  was reported \citep[e.g., see][]{2020ApJ...896..134P, 2022arXiv220413051R}. 

Being powerful emitters, blazars are frequently monitored in all the accessible wavelengths which resulted in accumulation of a substantial amount of data. The emission from blazars, predominantly of a nonthermal nature \citep[e.g.,][]{2017A&ARv..25....2P}, is dominated by Doppler-amplified emission from the jet, typically showing two broad peaks: the first at radio to X-ray bands, and the second at \grays. The low-energy component peak ($\nu^{\rm p}_{\rm syn}$) is used to further classify blazars as low- (LBL/LSP), intermediate- (IBL/ISP) or high- (HBL/HSP) peaked sources when $\nu^{\rm p}_{\rm syn}<10^{14}$ Hz, $10^{14}$ Hz $<\nu^{\rm p}_{\rm syn}<10^{15}$ and $\nu^{\rm p}_{\rm syn}>10^{15}$ Hz, respectively \citep{Padovani1995,Abdo_2010}. However, $\nu^{\rm p}_{\rm syn}$ can be well above $2.4\times10^{17}$ in extreme blazars \citep[e.g.,][]{sedentary,2001A&A...371..512C, Biteau2000, 1998ApJ...492L..17P, 2020MNRAS.496.5518S} which are interesting as they challenge our current understanding of particle acceleration and emission processes. In addition, the remarkable \gray\ blazar 4FGL J1544.3-0649, which was undetected in the X-ray and \grays\ until May 2017, showed a transient-like behaviour, becoming a very bright source for a few months and detected by Fermi Large Area telescope (\fermi) and MAXI X-ray sky monitor \citep{2021MNRAS.502..836S}. This suggests the existence of an undiscovered blazar population which may occasionally flare.

The broadband spectral energy distribution (SED) of blazars can be modeled within different scenarios involving the interaction of electrons and protons in a single or multiple zone. Although, there is a consensus that the low-energy component is due to the synchrotron emission of ultra-relativistic charged electrons in the blazar jet, the origin of the second component is highly debated. In the leptonic scenarios, this component is due to inverse Compton scattering of low-energy photons which might be the produced synchrotron photons \citep[synchrotron-self Compton model, SSC; ][]{1985A&A...146..204G, 1992ApJ...397L...5M, 1996ApJ...461..657B} or be of an external origin \citep[e.g.,][]{1994ApJ...421..153S}, such as photons directly emitted from the accretion disk \citep{1992A&A...256L..27D, 1994ApJS...90..945D} or reflected from the broad-line region \citep{1994ApJ...421..153S} or emitted from the dusty torus \citep{2000ApJ...545..107B}. Alternatively, the second component can be due to either the synchrotron emission of the energetic protons inside the jet \citep{2001APh....15..121M} or due to the secondaries generated in photo-pion and photo-pair interactions \citep{1993A&A...269...67M, 1989A&A...221..211M, 2001APh....15..121M, mucke2, 2013ApJ...768...54B, 2015MNRAS.447...36P, 2022MNRAS.509.2102G}. These hadronic models \citep[especially lepto-hadronic models, e.g.,][]{2018ApJ...863L..10A,2018ApJ...864...84K, 2018ApJ...865..124M, 2018MNRAS.480..192P, 2018ApJ...866..109S, 2019MNRAS.484.2067R,2019MNRAS.483L..12C, 2019A&A...622A.144S, 2019NatAs...3...88G, 2022MNRAS.509.2102G} have become more attractive after the observations of IceCube-170922A neutrino event from the direction of TXS 0506+056 \citep{2018Sci...361..147I, 2018Sci...361.1378I, 2018MNRAS.480..192P} as well as after the observations of multiple neutrino events from the direction of PKS 0735+178 when it was bright in the optical/UV, X-ray and \gray\ bands \citep{2022arXiv220405060S}.

Due to the extreme luminosities of blazars, even very high redshift ones can be observed \citep[e.g., see][]{2017ApJ...837L...5A}. The observation of distant blazars is of particular interest as they allow \text{i)} to study the relativistic jets as well as their connection with accretion disk/black hole in the early epochs of the Universe, \text{ii)} to measure the suppression of the \gray\ flux which can be used to estimate or constraint the density of the extragalactic background light (EBL) \citep{2004A&A...413..807K, 2007A&A...471..439M, 2008A&A...487..837F} and understand its cosmological evolution, \text{iii)} to investigate, in general, the properties of \gray\ emitting active galactic nuclei (AGN), which is important for the understanding of the cosmological evolution of the \gray\ background \citep{2010PhRvL.104j1101A}.

Due to their faintness, high redshift blazars are rather difficult to observe and identify, limiting the number of already associated high redshift blazars. For example, in the fourth catalog of AGNs detected by \fermi\ \citep[data release 3 (DR3);][]{2022ApJS..260...53A} only 110 blazars are observed beyond $z=2.0$ and only 10 beyond $z=3.0$. The most distant blazar observed in the \gray\ band is GB1508+5714 at $z=4.31$. The physical properties of these high redshift blazars have been frequently investigated using multi-frequency data \citep[e.g.,][]{2009MNRAS.399L..24G, 2011MNRAS.411..901G, 2015ApJ...804...74P, 2016ApJ...825...74P, 2017ApJ...839...96M, 2017ApJ...837L...5A, 2018ApJ...853..159L, 2019ApJ...871..211P}. For example, in \citet{2020ApJ...897..177P} by studying nine \gray\ emitting blazars and 133 candidate blazars with soft X-ray spectra it is shown that these high redshift blazars host massive black holes ($M_{\rm BH}>10^9\:{M_\odot}$) and have an accretion disk luminosity of $>10^{46}\:{\rm erg\:s^{-1}}$. Or, in \citet{2020MNRAS.498.2594S}, by studying the spectral and temporal properties of thirty-three distant blazars ($z>2.5$) and modeling their SEDs, it is found that the emission region size is $\leq0.05$ pc, while the magnetic field and the Doppler factor are correspondingly within $0.10-1.74$ G and $10.0-27.4$.

Although the number of observed high redshift blazars is not high enough to perform statistically sound population studies, the investigation of the properties of individual objects provides interesting peaces to understand the general physics of high redshift blazars. The multiwavelength monitoring of several high redshift blazars opens wide opportunities for investigation of their multiwavelength spectral and temporal properties as well as for performing detailed theoretical modeling and interpretation of the results. For example, the continuous monitoring of these sources in the HE \gray\ band by \fermi\ \citep{2009ApJ...697.1071A} allows to select various emission states, or their observations in the X-ray band with Neil Gehrels Swift Observatory \citep[][hereafter Swift]{2004ApJ...611.1005G}, and Nuclear Spectroscopic Telescope Array \citep[NuSTAR;][]{2013ApJ...770..103H} combined with the \gray\ data allows a precise estimation of the second emission component peak, or the data in the optical/UV bands can be used to constrain the high energy tail of the synchrotron component and/or the  direct thermal emission from the accretion disk \citep{2011MNRAS.411..901G}. Therefore, the data available in different bands can be used to put tighter constraints on the physics of individual high redshift blazars.

Here we present a broadband study of \source; at $z=3.10$ \citep{1978ApJ...226L..61W} it is one of the brightest high redshift blazars. It was observed in the X-ray band with various instruments (e.g., Einstein observatory \citep{1981ApJ...245..357Z}, ASCA \citep{1997ApJ...478..492C, 1996A&A...307....8S}, ROSAT \citep{1998ApJ...492...79F}, etc.) showing a particularly hard X-ray spectrum ($\sim1.2$). In the \gray\ band, with an energy flux of $(1.44\pm0.006)\times10^{-11}\:{\rm erg\:cm^{-2}\:s^{-1}}$ in the fourth catalog of \fermi\ AGNs \citep[DR3;][]{2022ApJS..260...53A}, it is the brightest blazar beyond $z=3.0$. Moreover, in several occasions \gray\ flares were observed when the daily flux was above $10^{-6}\:{\rm photon\:cm^{-2}\:s^{-1}}$ \citep{2022ATel15405....1V, 2020ATel14285....1A, 2017ATel10356....1C} which makes \source\ the most distant \gray\ flaring blazar \citep{2018ApJ...853..159L, 2020Ap....tmp...64S}.The broadband emission from \source\ was successfully modeled within a one-zone synchrotron and external inverse Compton scenario where the excess in optical/UV band was interpreted as emission from bright thermal accretion disk \citep{2010A&A...509A..69B}.

In general, the peak of the second component in the SED of high redshift blazars is at MeV energies, which implies their HE \gray\ spectrum is soft, so they are not ideal sources for \gray\ observations. Therefore, the observation of the \gray\ flaring activity of distant blazars, which is crucial for testing different emission scenarios of relativistic jets, is even more interesting as compared with that of the nearby sources. Motivated \textit{i)} by the availability of multiwavelength data from \source\ observations - since 2008 in the HE \gray\ band by \fermi, multiple observations of \source\ by Swift X-Ray Telescope (XRT) and Ultra-violet Optical Telescope (UVOT) instruments and two observations of \source\ with NuSTAR, and \textit{ii)} by the observed multiple flaring activities of \source, we decided to investigate the spectral and temporal properties of \source\ by analyzing the data accumulated in the optical/UV, X-ray and \gray\ bands and put, through theoretical modeling, a constraint on the physical processes responsible for the \source\ emission in the quiescent and flaring states. 

The paper is structured as follows. The data extraction and analysis in the \gray, X-ray and optical/UV bands are presented correspondingly in Sections \ref{data_anal}, \ref{nustar} and \ref{swift}. The SED of \source\ and its evolution in time is presented in Section \ref{temp_seds}, and the origin of the emission is discussed in Section \ref{org}. The results are presented and discussed in section \ref{res} while the summary is given in Section \ref{summary}.

\section{\fermi\ observations and data analyses}\label{data_anal}
Fermi satellite launched in 2008 carries two instruments- the Large Area Telescope (LAT) is the main instrument on board designed to scan the entire sky in \gray\ band, and the Gamma-ray Burst Monitor (GBM) is designed to study gamma-ray bursts. LAT is a pair-conversion \gray\ telescope sensitive in the energy range from 20 MeV to 300 GeV with a field of view of $\sim2.4$ sr. It is by default in the all sky scanning mode which allows to study the HE properties of various sources, including blazars. For more details on the LAT see \citet{2009ApJ...697.1071A}.

We have analyzed the \gray\ data collected between August 4 2008 and September 9 2022 (MET=239557417–686130659). The data was reduced and analyzed following the standard procedures described in the \fermi\ documentation with \textit{fermitools} version 2.0.8 using P8R3\_SOURCE\_V3 instrument response functions. The events in the energy range from 100 MeV to 300 GeV are selected from a circular region of interest (ROI) of $12^\circ$ radius centered at the \gray\ location of \source\ (RA$=89.99$, Dec$=-28.65$), retrieving only events classified as {\it evclass=128} and {\it evtype= 3}. A zenith angle cut less than $90^\circ$ was introduced to remove the secondary \grays\ from the earth limb. The model file that includes point-like sources and background models was created based on the \fermi\ fourth source catalog (4FGL) incremental version \cite[DR 3;][]{2022ApJS..260...53A}, which is based on 12 years of initial \fermi\ operation and includes best-fit spectral parameters of all known 4FGL \gray\ sources in the ROI. The sources which are within $17^\circ$ from the \source\ location were included in the model file; the spectral parameters of the sources within $12^\circ-17^\circ$ are fixed to their values reported in 4FGL while they are left free for the sources falling within $<12^\circ$ radius. The galactic background and isotropic galactic emissions were modeled with the latest version available files, gll\_iem\_v07 and iso\_ P8R3\_SOURCE\_V3\_v1, respectively. 

The \gray\ analysis is performed with {\it gtlike} tool, following the binned likelihood method. Initially, the spectrum of \source\ was modeled with a log-parabolic model as in 4FGL. However, the fit was also performed when assuming a power-law model for \source\ \gray\ emission and the resulting model file was used in the light-curve calculations, because for shorter periods a power-law can be a good approximation of the spectrum. The significance of the source emission is estimated using test statistic (TS), which is defined by $TS = {\rm 2(lnL_1-lnL_0)}$ where $L_1$ and $L_0$ are maximum likelihoods with and without the source, respectively \citep{1996ApJ...461..396M}.

In order to investigate the variability pattern of the source, the light curves were generated by two different methods. Initially, the whole time interval was divided into 5-day intervals and the photon index and flux of \source\ were estimated by the unbinned analysis method from {\it gtlike} tool. Next, in order to obtain a deeper insight into the \gray\ flux evolution in time, the adaptively binned light curve was computed \citep{2012A&A...544A...6L}. In this method, the bin widths above the optimal energy ($E_{\rm opt}$) are adjusted to have fixed uncertainty, so in the case of flux increase shorter intervals are estimated, whereas in the quiescent/normal states time bins are wider. This method has been proven to be a powerful tool in finding flaring periods in blazars \citep[e.g., see][]{2013A&A...557A..71R, 2016ApJ...830..162B, 2017MNRAS.470.2861S, 2017A&A...608A..37Z,2017ApJ...848..111B,  2018ApJ...863..114G, 2018A&A...614A...6S, 2021MNRAS.504.5074S, 2022MNRAS.517.2757S, 2022MNRAS.513.4645S}.

\begin{figure*}
	\includegraphics[width=\textwidth, height=0.9\textheight]{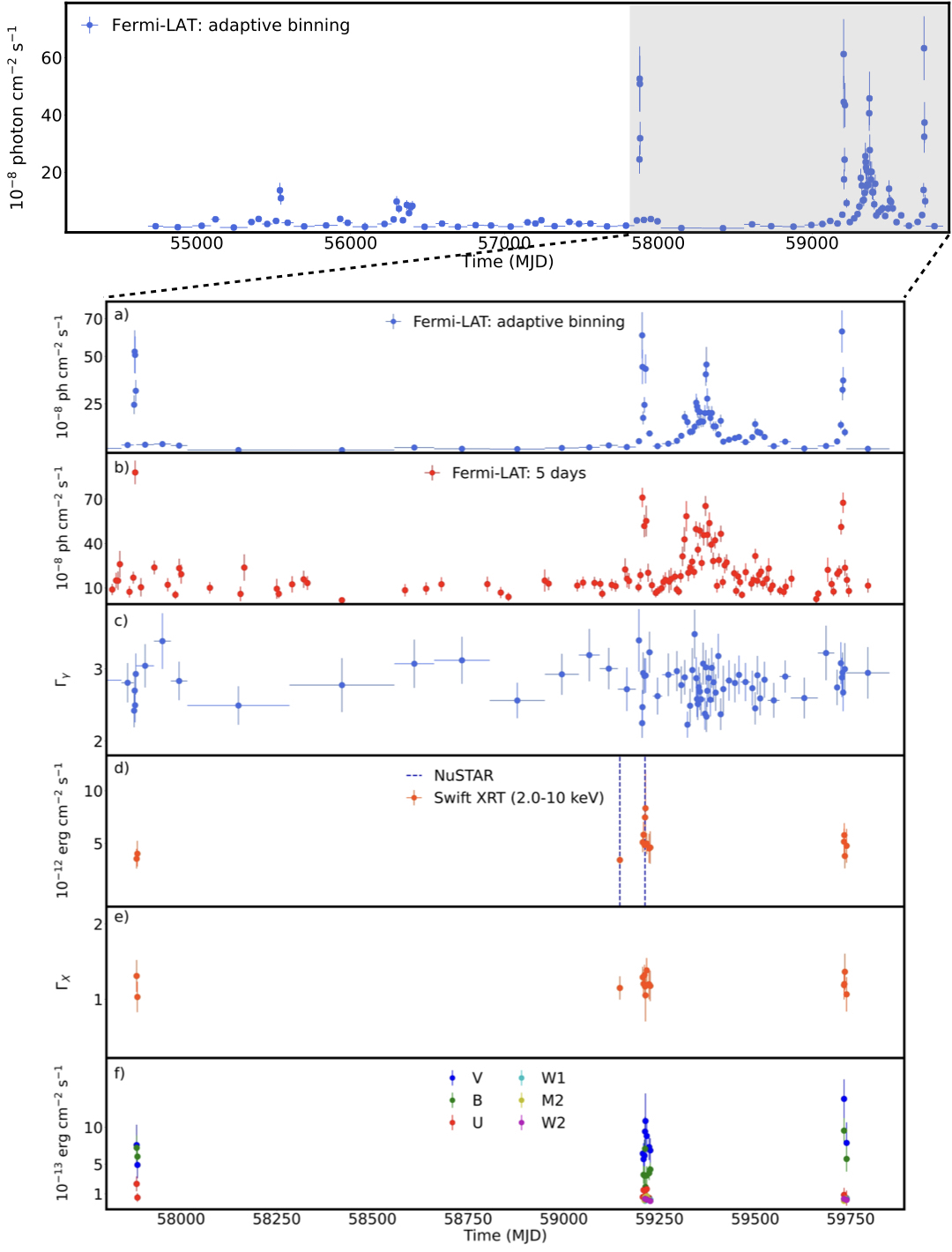}
    \caption{Multiwavelength light curve of \source. Top panel shows the long-term adaptively binned \gray\ light curve above 168.19 MeV. The other panels show the light curves after MJD 57800 (16 February 2017) when the source was active in the \gray\ band. \textit{a)} adaptively binned \gray\ light curve, \textit{b)} 5-day binned \gray\ light curve ($>100$ MeV), \textit{c)} \gray\ photon index measured for the adaptive time bins, \textit{d)} and \textit{e)} X-ray flux (2-10 keV) and photon index in different Swift observations. The dashed blue lines show the NuSTAR observation times. \textit{f)} Swift UVOT fluxes in V, B, U, W1, M2, and W2 bands.}
    \label{mw_lc}
\end{figure*}

The adaptively binned light curve ($>E_{\rm opt}=168.19$ MeV) is shown in Fig. \ref{mw_lc} upper panel. Up to MJD 57740 the \gray\ flux was in its average level of $(1-3)\times10^{-8}\:{\rm photon\:cm^{-2}\:s^{-1}}$ with no significant changes, while then, in several occasions, the \gray\ flux increased substantially. The light curve with 5-day ($>100$ MeV) and adaptive bins ($>E_{\rm opt}=168.19$ MeV) for the period when the source was active in the \gray\ band are shown correspondingly in Fig. \ref{mw_lc} panels a) and b). The first flaring period was between MJD 57876-57883 when the flux increased with a maximum of $(5.26\pm1.13)\times10^{-7}\:{\rm photon\:cm^{-2}\:s^{-1}}$. Starting from MJD 59170, the source entered an active emission state with several bright flaring periods between MJD 59204-59233, MJD 59301-59411 and MJD 59721-59738. The maximum \gray\ flux of the source, $(6.32\pm1.11)\times10^{-7}\:{\rm photon\:cm^{-2}\:s^{-1}}$ was also observed in these \gray\ flaring periods.

Fig. \ref{mw_lc} panel c shows the \gray\ photon index estimated for the adaptively binned periods; it varies in time as well. In the non-flaring periods, the \gray\ spectrum is characterised by a soft spectrum with a mean of  $\Gamma\simeq2.83$ but the photon index hardens during the bright periods as can be seen from Fig. \ref{mw_lc} panel c. For example, during the first flare between MJD 57876-57883 the hardest index of $2.49\pm0.23$ was observed on MJD 57879.9 or during the second flare between MJD 59204-59233 the hardest index was $2.25\pm0.21$ when the source was in an active state with a flux of $(6.12\pm1.22)\times10^{-7}\:{\rm photon\:cm^{-2}\:s^{-1}}$. During the hardest \gray\ emission period, $2.23\pm0.18$ was detected on MJD 59322 which is unusual for this source.  

\section{NuSTAR data analysis}\label{nustar}
NuSTAR is a hard X-ray telescope sensitive in the 3-79 keV energy range \citep{2013ApJ...770..103H}. NuSTAR with two focal plane modules (FPMs), FPMA and FPMB, observed \source\ on 28 December 2020 (MJD 59211.99) and on 24 October 2020 (MJD 59146.17) for 97.1 ks and 24.3 ks, respectively. It should be noted that around these observations \source\ was also monitored with Swift (see dashed blue lines in Fig. \ref{mw_lc} panel d), so the X-ray spectrum of the source can be obtained in a large range of 0.3-79 keV.

The NuSTAR data was analyzed applying the standard procedure and using \textit{NuSTAR\_Spectra} tool developed in \citet{2022MNRAS.514.3179M}. \textit{NuSTAR\_Spectra} script downloads calibrated and filtered event files from the SSDC repository, uses 
\textit{XIMAGE} package to precisely locate the source's coordinate then extracts high-level scientific products for the detected sources using {\it nuproducts} routine. The script automatically sets the source extraction region radius depending on the source counts (usually in the range of 30-70 arcsec). The background is computed in an annulus centered on the source with a minimum separation of 50 arcsec between the inner and outer radii. Then, a spectral analysis is performed using the XSPEC package \citep{1996ASPC..101...17A} adopting Cash statistics \citep{1979ApJ...228..939C}. More details on \textit{NuSTAR\_Spectra} are available in \citet{2022MNRAS.514.3179M}.

The analysis shows that the X-ray photon index of \source\ is the same in both observations - $1.26\pm0.06$ and $1.26\pm0.02$ on MJD 59146.17 and MJD 59211.99, respectively. The X-ray flux between 3-10 keV measured on MJD 59146.17 is ${\rm F_{3-10\:keV}}=(2.72\pm0.06)\times10^{-12}\:{\rm erg\:cm^{-2}\:s^{-1}}$ and on MJD 59211.99, it increased by about a factor of two, ${\rm F_{3-10\:keV}}=(5.10\pm0.04)\times10^{-12}\:{\rm erg\:cm^{-2}\:s^{-1}}$. Similarly, the flux between 10-30 keV also increased in these two observations, being correspondingly ${\rm F_{10-30\:keV}}=(5.79\pm0.20)\times10^{-12}\:{\rm erg\:cm^{-2}\:s^{-1}}$ and ${\rm F_{10-30\:keV}}=(1.08\pm0.01)\times10^{-11}\:{\rm erg\:cm^{-2}\:s^{-1}}$. This shows that the source in the 3.0-30 keV range was in an enhanced state on 28 December 2020. 
\section{Swift data analysis}\label{swift}
Swift is a space-based observatory with three main instruments onboard, namely burst alert telescope (BAT) sensitive in the energy range of 3.0-150.0 keV, XRT sensitive in the energy range of 0.3-10.0 keV, and UVOT sensitive in the optical/UV band 170 - 650 nm \citep{2004ApJ...611.1005G}. Swift performed 29 observations of \source\ among which nine observations were performed before the lunch of \fermi. However, in order to investigate the flux changes in different years, we have analyzed all the data from Swift observations of \source.
\subsection{Swift XRT}
All the XRT observations were processed with {\it Swift\_xrtproc} tool applying standard analysis procedure \citep{2021MNRAS.507.5690G}. {\it Swift\_xrtproc} downloads the raw data for each snapshot and for the whole observation, generates exposure maps and calibrated data product using the XRTPIPELINE task adopting standard parameters and filtering criteria. The source region counts are estimated from a circle of a radius of 20 pixels  while the background counts from an annular region centred around the source with a radius sufficiently large to avoid contamination from source photons. The resultant ungrouped data is loaded in XSPEC \citep{1996ASPC..101...17A} for spectral fitting using Cash statistics \citep{1979ApJ...228..939C}, modeling the source spectrum as power-law and log-parabola. As a result, the X-ray photon index in the energy range 0.3-10 keV and the flux in various bands are estimated.

The 2-10 keV X-ray flux variation is shown in Fig. \ref{mw_lc} panel d). Although in the X-ray band there is a limited number of observations, the flux variation is evident. The X-ray emission of the source in the 2.0-10 keV band was at the level of $\sim3.0\times10^{-12}\:{\rm erg\:cm^{-2}\:s^{-1}}$ but during the bright periods it is $\geq5.0\times10^{-12}\:{\rm erg\:cm^{-2}\:s^{-1}}$. The highest X-ray flux of $(8.34\pm3.59)\times10^{-12}\:{\rm erg\:cm^{-2}\:s^{-1}}$ was observed on MJD 59213.18. The X-ray spectrum of the source is hard (Fig. \ref{mw_lc} panel e) and during all the observations $\Gamma_{\rm X-ray}\leq1.38$. Therefore, as it is typical for FSRQs, the X-ray band defines the rising part of the second component in the SED.   
\subsection{Swift UVOT}
In the same periods, UVOT observed \source\ in V (500-600 nm), B (380- 500 nm), U (300- 400 nm), W1 (220-400 nm), M2 (200-280 nm) and W2 (180–260 nm) filters. All the available 28 observations were downloaded and reduced using HEAsoft version 6.29 with the latest release of HEASARC CALDB. Photometry was computed using a five-arcsecond source region centered on the sky position of \source\ and the background counts are estimated from a twenty-arcsecond region away from the source. The magnitudes were derived using {\it uvotsource} tool, then the fluxes were obtained using the conversion factors provided by \citet{2008MNRAS.383..627P} which were corrected for extinction using the reddening coefficient $E(B-V)$ from the Infrared Science Archive \footnote{http://irsa.ipac.caltech.edu/applications/DUST/}.

Fig. \ref{mw_lc} panel f) shows the light curve of \source\ in optical/UV bands. The source is relatively faint in all the filters with the flux around $\simeq10^{-13}\:{\rm erg\:cm^{-2}\;s^{-1}}$. In some cases, coinciding with the flares in the \gray\ band, the flux increased several times. The highest flux of the source was observed in V-band; on MJD 59213.18 and MJD 59732.67 it was $(1.08\pm0.37)\times10^{-12}\:{\rm erg\:cm^{-2}\;s^{-1}}$ and $(1.38\pm0.26)\times10^{-12}\:{\rm erg\:cm^{-2}\;s^{-1}}$, respectively. In addition, VOU-Blazar tool, which allows to search and collect all spectral information accessible through virtual observatory services and build the multiwavelength SEDs of blazars \citep{2020A&C....3000350C} was used to investigate the source emission properties in the infrared band. In particular, data extracted from the Wide-field Infrared Survey Explorer (WISE) and NEOWISE surveys \citep{2014ApJ...792...30M} show that the source emission at 3.4 and 4.6 $\mu m$ wavelengths (infrared) was at the level of several times $10^{-13}\:{\rm erg\:cm^{-2}\;s^{-1}}$.
\section{Multiwavelength SEDs}\label{temp_seds}
The data analyzed in this paper allows to build the SEDs of \source\ in different periods. The single snapshot SED provides substantial information on the source emission properties whereas the variation of these SEDs in time is crucial for understanding the dynamical changes in the emission components. For this purpose, we generated SED/Light curve animation of \source\ by showing the \gray\ spectra with all available data sets. For each adaptively binned interval we performed \gray\ spectral analysis using the unbinned likelihood method implemented in {\it gtlike} tool. Then, for each \gray\ period, together with the \gray\ data we plotted the Swift XRT, NuSTAR and Swift UVOT data as well as archival data extracted with VOU-blazar tool. By going from one to another \gray\ period we can investigate the changes in the multiwavelength SED of \source.

The SED/light curve animation is available here \href{https://youtu.be/4UPqf-C7EWc}{\nolinkurl{youtube.com/4UPqf-C7EWc}} . As the blazar is at $z=3.10$ the UVOT flux could be affected by absorption of neutral hydrogen in intervening Lyman-$\alpha$ absorption systems \citep[e.g.,][]{2011MNRAS.411..901G} which was corrected using the attenuation calculated in \citet{2010MNRAS.405..387G} for the UVOT filters. The SED/light curve animation shows the high amplitude changes observed in the \gray\ band; the gray background data points show the \gray\ flux estimated in different periods. Also, the spectral hardening in several bright \gray\ periods can be seen.  
\begin{figure*}
	\includegraphics[width=0.48\textwidth]{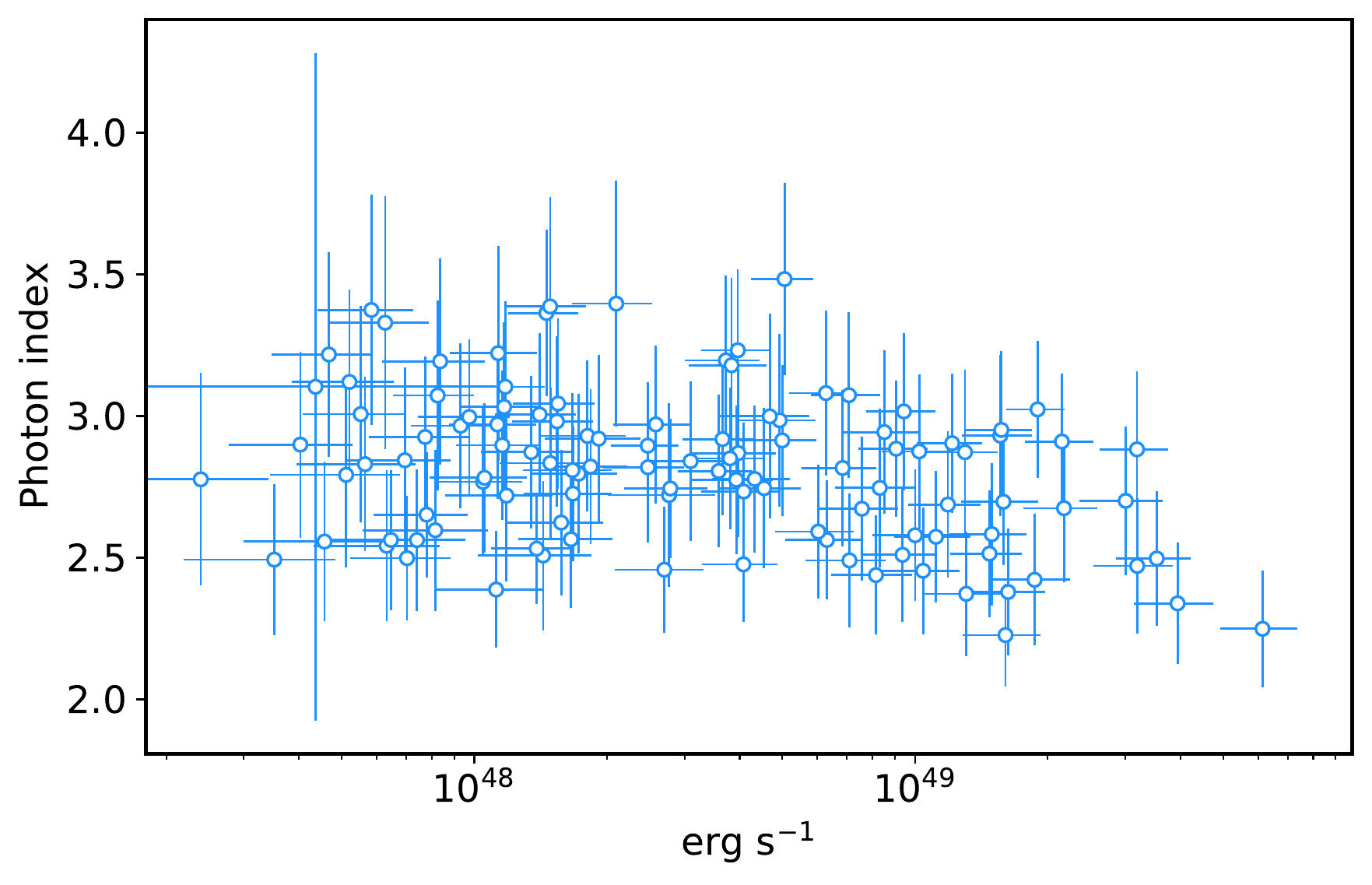}
	\includegraphics[width=0.48\textwidth]{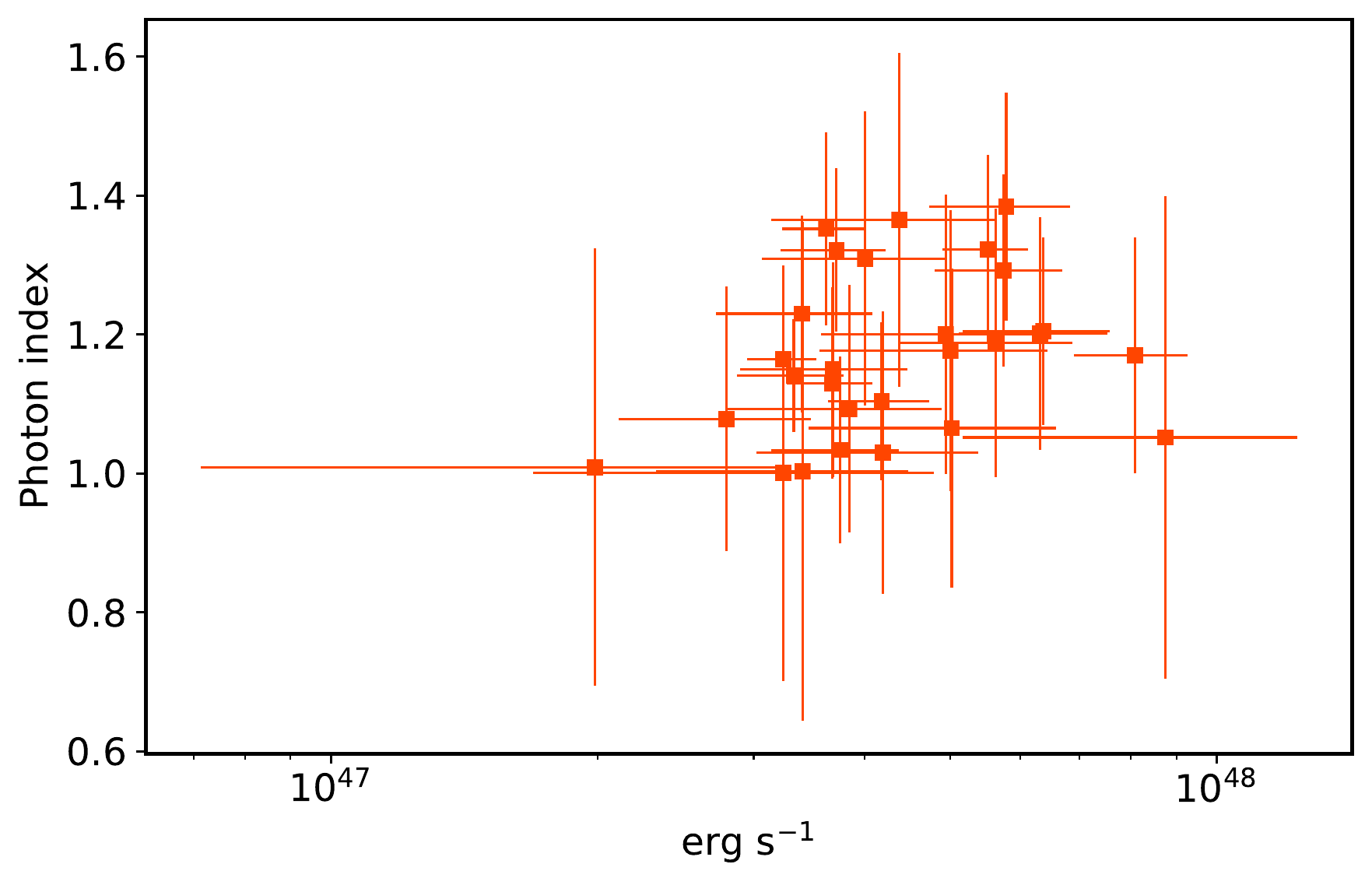}
    \caption{The luminosity versus the spectral index in the \gray\ (left panel) and X-ray bands (right panel).}
    \label{lum_index}
\end{figure*}
\section{Origin of broadband emission}\label{org}
In the previous section, the generated multiwavelength SEDs show the features of \source\ emission in different periods. It is especially important to investigate the processes taking place in the jet of high redshift blazars as they could provide information on the jet plasma state in the early Universe. For this reason, the following periods were considered for modeling:
\begin{itemize}
  \item The SED between MJD 55150-55330 when the source was in the quiescent state, i.e., the \gray\ flux above 100 MeV was $(2.77\pm0.84)\times10^{-8}\:{\rm photon\:cm^{-2}\:s^{-1}}$ and the 2-10 keV X-ray flux  was $(3.29\pm1.11)\times10^{-12}\:{\rm erg\:cm^{-2}\:s^{-1}}$.
  \item The SED between MJD 59208-59212 when the source was bright in the \gray\ and X-ray (2-10 keV) bands with corresponding fluxes of $(5.46\pm0.83)\times10^{-8}\:{\rm photon\:cm^{-2}\:s^{-1}}$ and $(7.47\pm1.18)\times10^{-12}\:{\rm erg\:cm^{-2}\:s^{-1}}$. This period coincides with the NuSTAR observation on MJD 59211.99 which showed that the source was in an elevated X-ray emission state also in the 3-30 keV range.   
\end{itemize}

The broadband SEDs were modeled using a one-zone leptonic scenario. In this model, it is assumed that the emission originates from a spherical blob of radius $R$ moving with a bulk Lorentz factor of $\Gamma$ at a small inclination angle of $\theta$ with respect to the observer. Due to the relativistic motion and small $\theta$ the radiation will be Doppler amplified by a factor of $\delta\simeq\Gamma$. The emission region magnetized with a field strength of $B$ is filled with relativistic electrons whose energy distribution is given by
\begin{equation}
    N_{\rm e}=N_0\:\gamma^{-p}\:exp(-\frac{\gamma}{\gamma_{\rm cut}})\:\:\:\:\:\:\:\:\:\: \gamma>\gamma_{\rm min}
    \label{el_spec}
\end{equation}
where $\gamma=E_{\rm e}/m_ec^2$ is the electron Lorentz factor, $p$ is the power-law index, $\gamma_{\rm min}$ and $\gamma_{\rm cut}$ are the minimum and cut-off energy, respectively. The parameter $N_0$ defines the electron energy density through $U_{\rm e}=m_e c^2\int\gamma N_e(\gamma)d\gamma$.

The electrons with energy distribution given by Eq. \ref{el_spec} undergo synchrotron losses under the magnetic field, producing the data observed between radio to X-ray bands. Instead, the second component in the SED, from X-rays to \grays, is from the inverse Compton scattering of internal and external photons on the same population of the electrons. When the electrons upscatter the synchrotron photons, the second component is explained by the SSC component \citep{1985A&A...146..204G, 1992ApJ...397L...5M, 1996ApJ...461..657B}. Alternatively, if the emission region is within the BLR, the second component can be due to external Compton scattering of direct disk radiation \cite[EIC-disk;][]{1992A&A...256L..27D, 1994ApJS...90..945D} and/or due to external Compton scattering of photon reflected from BLR clouds \citep[EIC-BLR;][]{1994ApJ...421..153S}. Instead, if the jet energy dissipation occurs at larger distances it can be due to external Compton scattering of dusty torus photons \citep[EIC-torus;][]{2000ApJ...545..107B}. 

In this paper, for a general view we consider three different scenarios: \textit{i)} the broadband emission from \source\ is entirely due to synchrotron/SSC radiation, \textit{ii)} the jet dissipation region is close to the central black hole, and SSC, EIC-disk and EIC-BLR are contributing to the HE component and \text{iii)} the emission region is beyond the BLR and the HE component is due to EIC-torus. It is assumed that BLR is a spherical shell \citep[e.g.,][]{2003APh....18..377D} with lower and upper boundaries of $0.9\times R_{\rm BLR}$ and $1.2\times R_{\rm BLR}$, respectively. $R_{\rm BLR}$ is assumed to scale as $R_{\rm BLR}=10^{17}\:L_{\rm disc, 45}^{0.5}$ cm where $L_{\rm disc, 45}=L_{\rm disc}/10^{45}\:{\rm erg\: s^{-1}}$ is the accretion disk luminosity \citep{2015MNRAS.448.1060G}. Similarly, we assume that the distance of dusty torus is $2\times10^{18}\:L_{\rm disc, 45}^{0.5}$ \citep{2015MNRAS.448.1060G} which emits $\eta=0.5$ fraction of disk luminosity in the IR range for which we adopted $T_{\rm torus} = 10^3$ K effective temperature. The disk luminosity and effective temperature are correspondingly $8.7\times10^{46}\:{\rm erg\:s^{-1}}$ and $T_{\rm disk} = 1.9\times10^4$ K estimated by fitting the thermal blue-bump component in the SED with a black-body component.

The remaining free model parameters are $p$, $\gamma_{\rm min}$, $\gamma_{\rm cut}$, $U_{\rm e}$, $B$ and $R$ which should be constrained during the fitting. The SED fitting is performed using publicly available code JetSet which is a numerical code allowing to fit the radiative models to data and obtain the parameters statistically better explaining them \citep{2006A&A...448..861M, 2009A&A...501..879T, 2011ApJ...739...66T,2020ascl.soft09001T}. These parameters are initially constrained by using the Minuit optimizer and then improved by Markov Chain Monte Carlo (MCMC) sampling of their distributions. We applied the EBL  model from \citet{2008A&A...487..837F} to correct the attenuation in the HE \gray\ band, but as the \gray\ data extends to several tens of GeV it affects only the model extrapolation to higher energies.
\section{Results and Discussions}\label{res}
At $z=3.10$, \source\ is one of the most powerful FSRQs in the extragalactic \gray\ sky; the time-averaged \gray\ luminosity of the source is $1.90\times10^{48}\:{\rm erg\:s^{-1}}$ (assuming a distance of $27.08$ Gpc). However, in several occasions, the source shows bright \gray\ flares when the flux substantially increases and the spectrum hardens. Fig. \ref{lum_index} left panel shows the \gray\ luminosity of \source\ versus the photon index. During the bright periods, the luminosity increases, being above $10^{49}\:{\rm erg\:s^{-1}}$; the maximum \gray\ luminosity corresponds to $6.14\times10^{49}\:{\rm erg\:s^{-1}}$. It should be noted that among 113 adaptively binned intervals, the source luminosity was above $10^{49}\:{\rm erg\:s^{-1}}$ in 25 intervals amounting $61.8$ days when extreme \gray\ luminosity was observed. Photon index hardening with increasing luminosity/flux can be noticed in Fig. \ref{lum_index} left panel. In order to test possible correlation/anti-correlation between the luminosity and photon index, a Pearson correlation test was applied which yielded $-0.39$ with a probability of $P=1.2\times10^{-5}$. This indicates moderate anti-correlation between the luminosity and photon index, that is when the source emission becomes brighter the photon index hardens (harder-when-brighter trend). It should be noted that for blazars such trend is frequently observed in different bands \citep[e.g.,][]{2010ApJ...710..810A, 2010ApJ...721.1425A, 2017ApJ...848..111B, 2021MNRAS.502..836S, 2021MNRAS.504.5074S, 2018ApJ...863..114G, 2022MNRAS.513.4645S} which can be interpreted as interplay between acceleration and cooling of the electrons \citep{1998A&A...333..452K}.

\source\ shows also interesting features in the X-ray band. The X-ray photon index versus the $0.3-10$ keV X-ray luminosity is shown in Fig. \ref{lum_index} right panel. The X-ray emission is characterized by a hard spectrum ($\Gamma_{\rm X-ray}<1.38$) with a high luminosity ($>10^{47}\:{\rm erg\:s^{-1}}$). It should be noted that \textit{XMM-Newton} observations of \source\ also showed a high luminosity of $2\times10^{47}\:{\rm erg\:s^{-1}}$ with a spectral index of $1.27\pm0.02$ \citep{2001A&A...365L.116R}. There is no evidence of softening or hardening when the source gets brighter in the X-ray band; the highest luminosity in the X-ray band is $8.74\times10^{47}\:{\rm erg\:s^{-1}}$ observed on MJD 59213.18. Similarly, the $3-30$ keV X-ray luminosity was $1.40\times10^{48}\:{\rm erg\:s^{-1}}$ on MJD 59211.99 \ and $7.47\times10^{47}\:{\rm erg\:s^{-1}}$ on MJD 59146.17.

The SED of \source\ was assembled in the flaring and quiescent periods (see Fig. \ref{SED}). Comparing and contrasting the jet parameters obtained through modeling of the SED in different periods is crucial, allowing to understand the processes at work in the jet of \source.
\subsection{Synchrotron/SSC emission from the jet}
\begin{figure*}
	\includegraphics[width=0.48\textwidth]{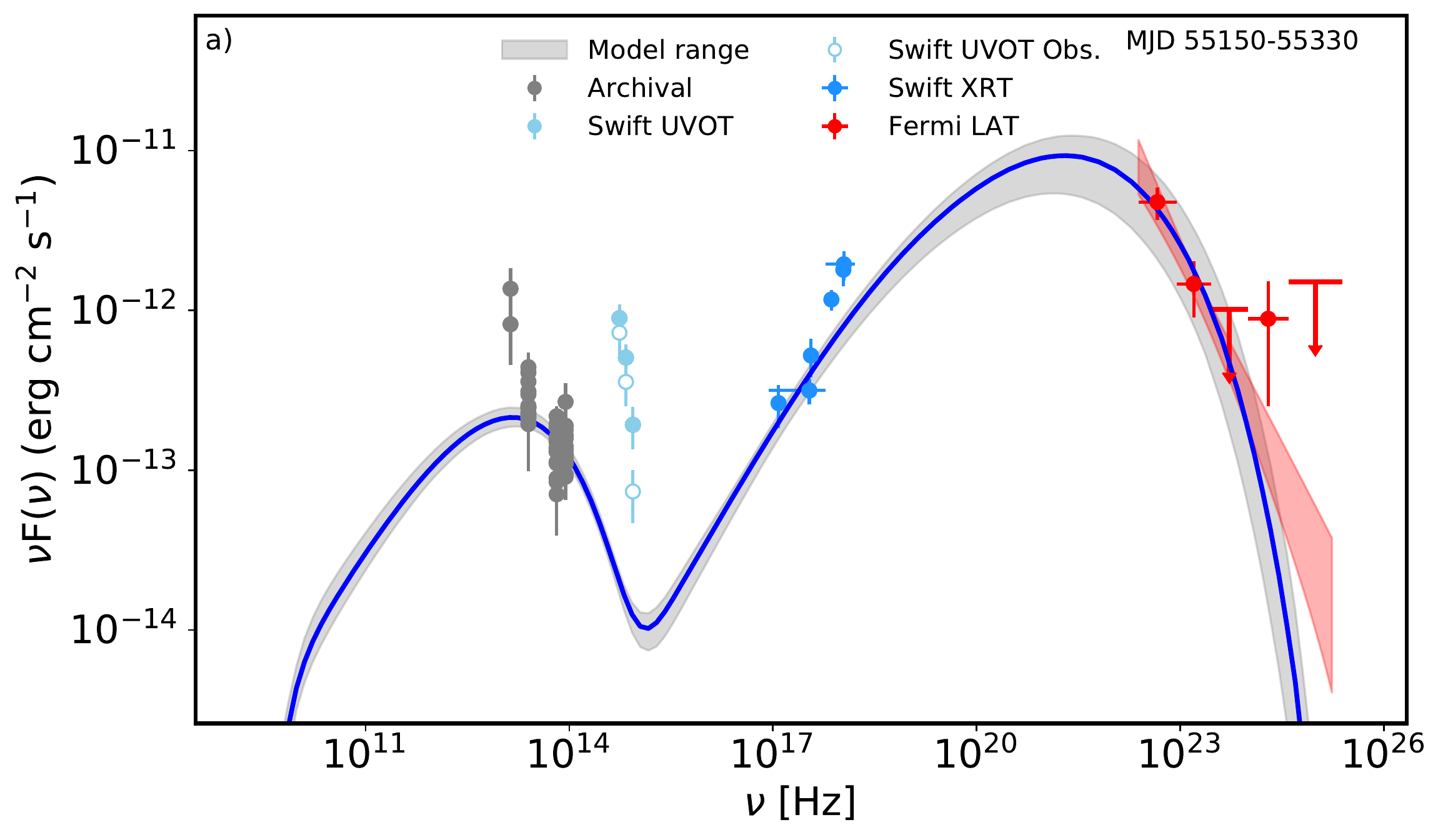}
	\includegraphics[width=0.48\textwidth]{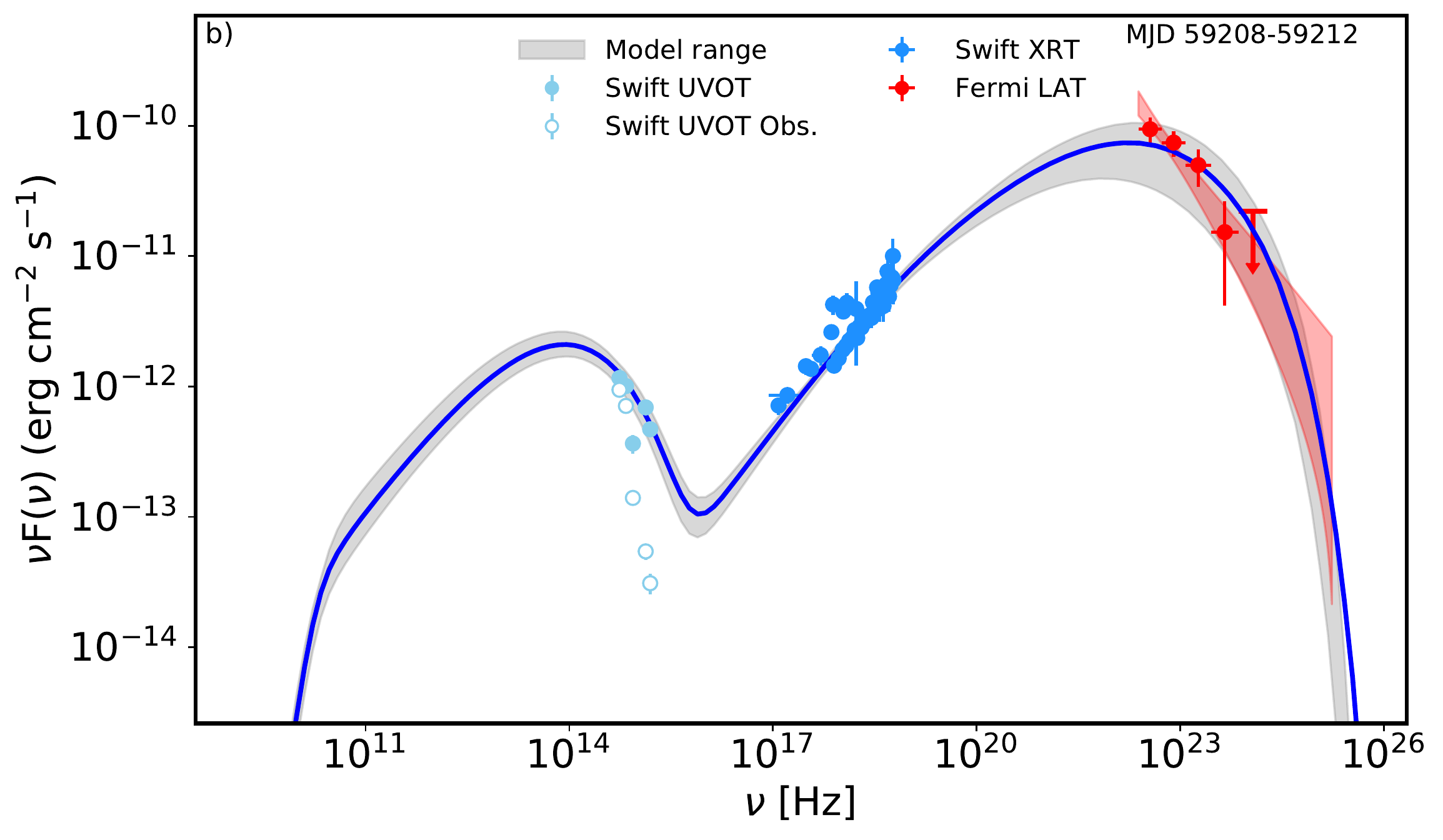}\\
	\includegraphics[width=0.48\textwidth]{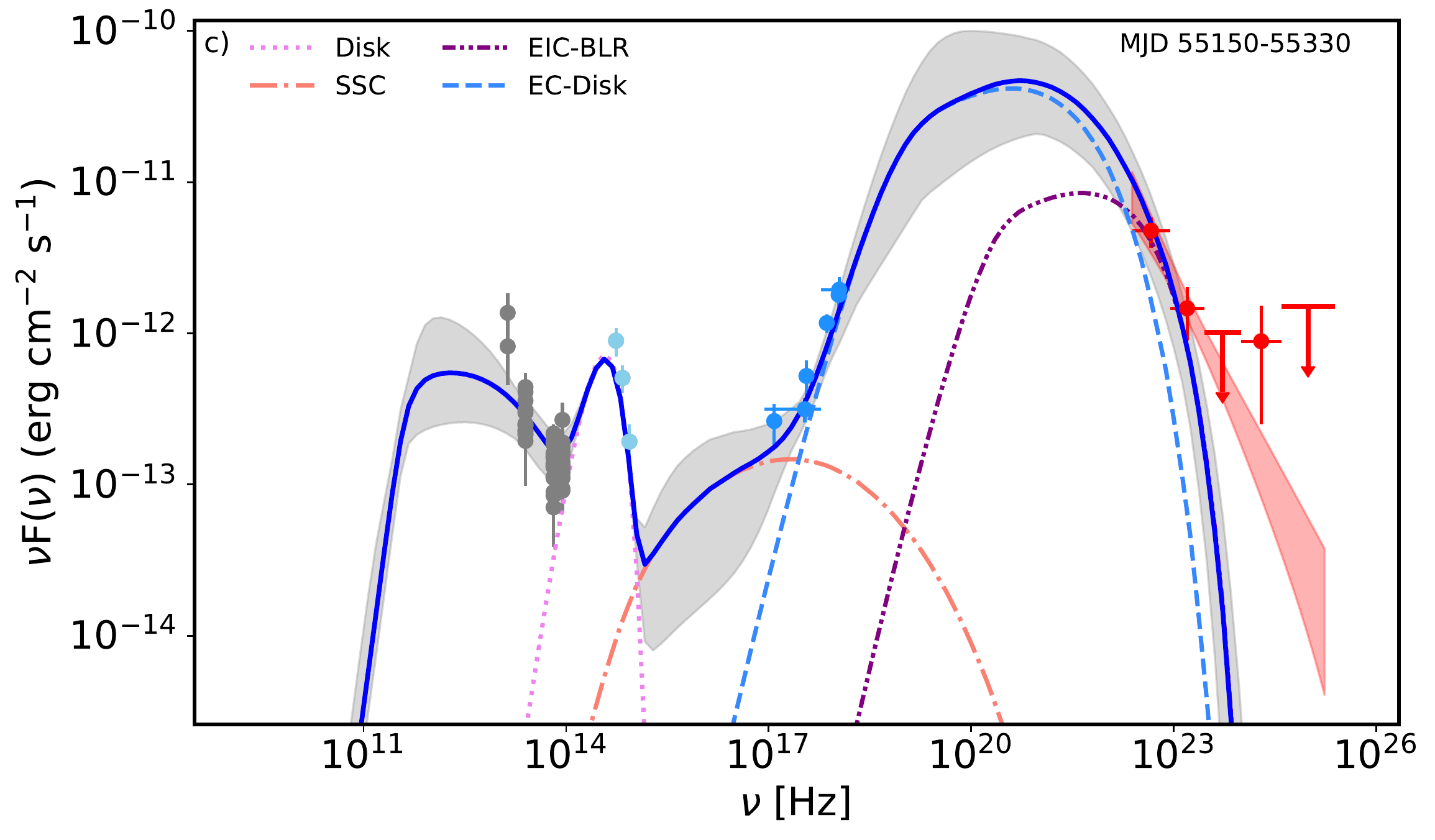}
	\includegraphics[width=0.48\textwidth]{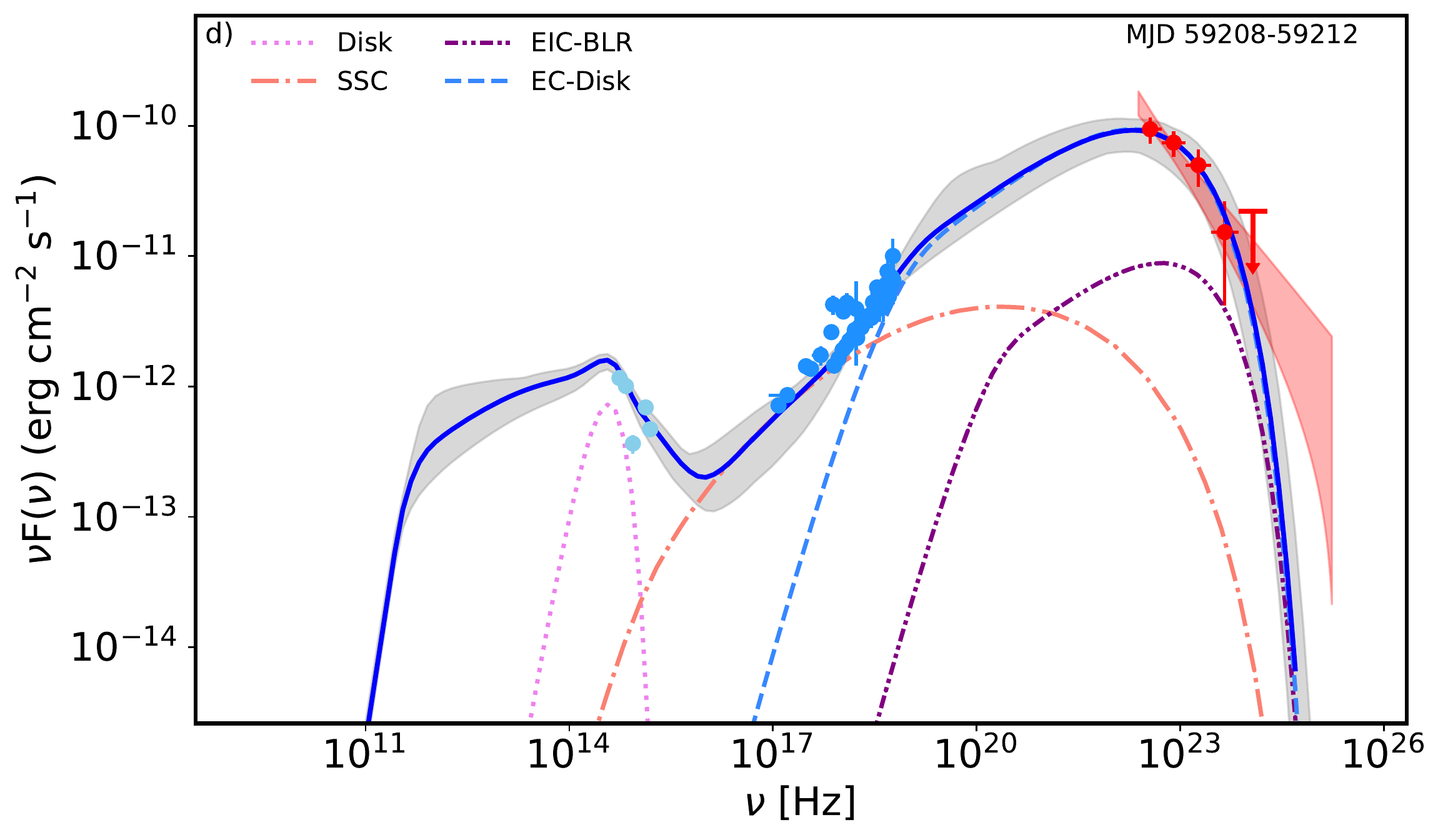}\\
	\includegraphics[width=0.48\textwidth]{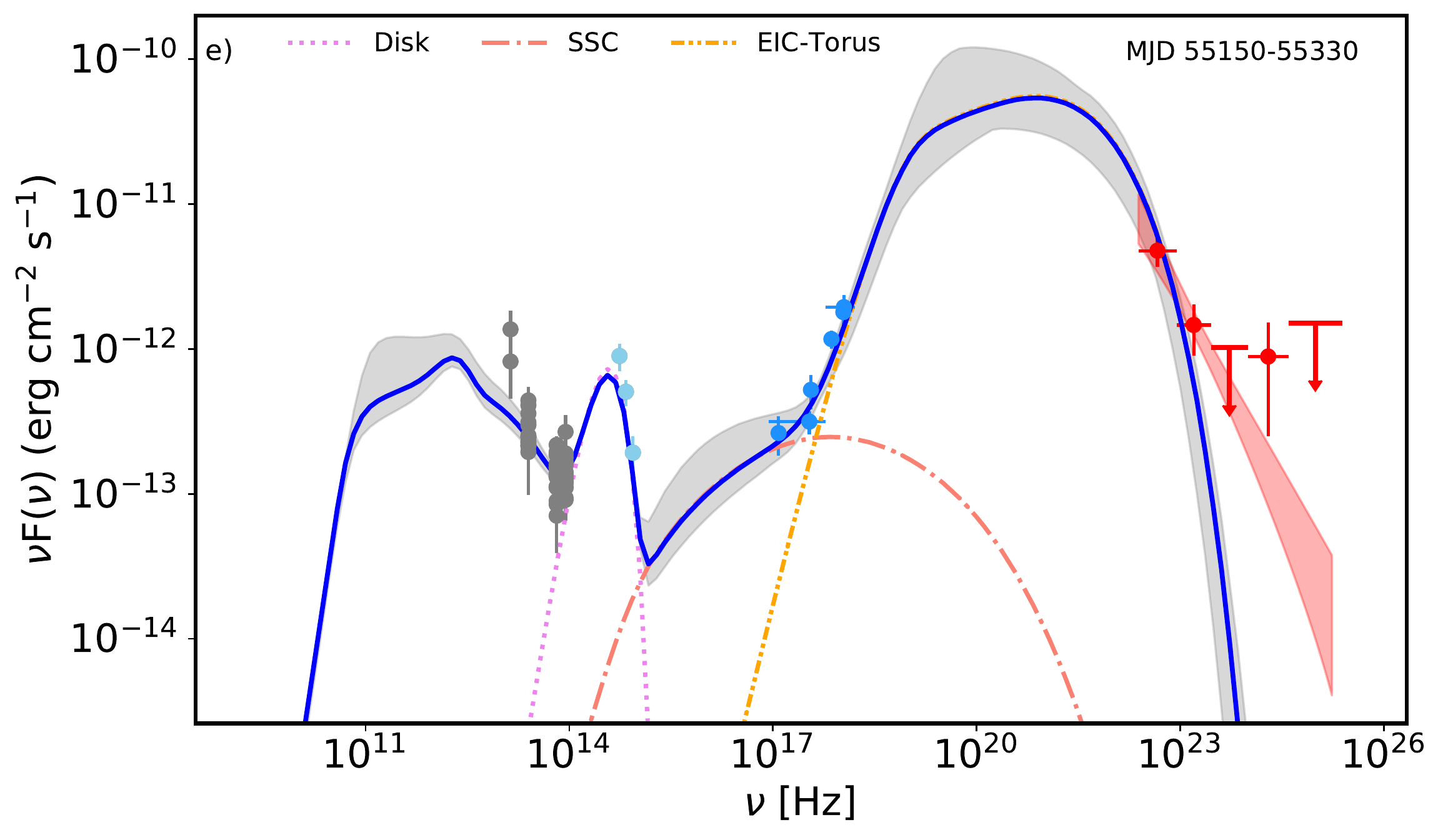}
	\includegraphics[width=0.48\textwidth]{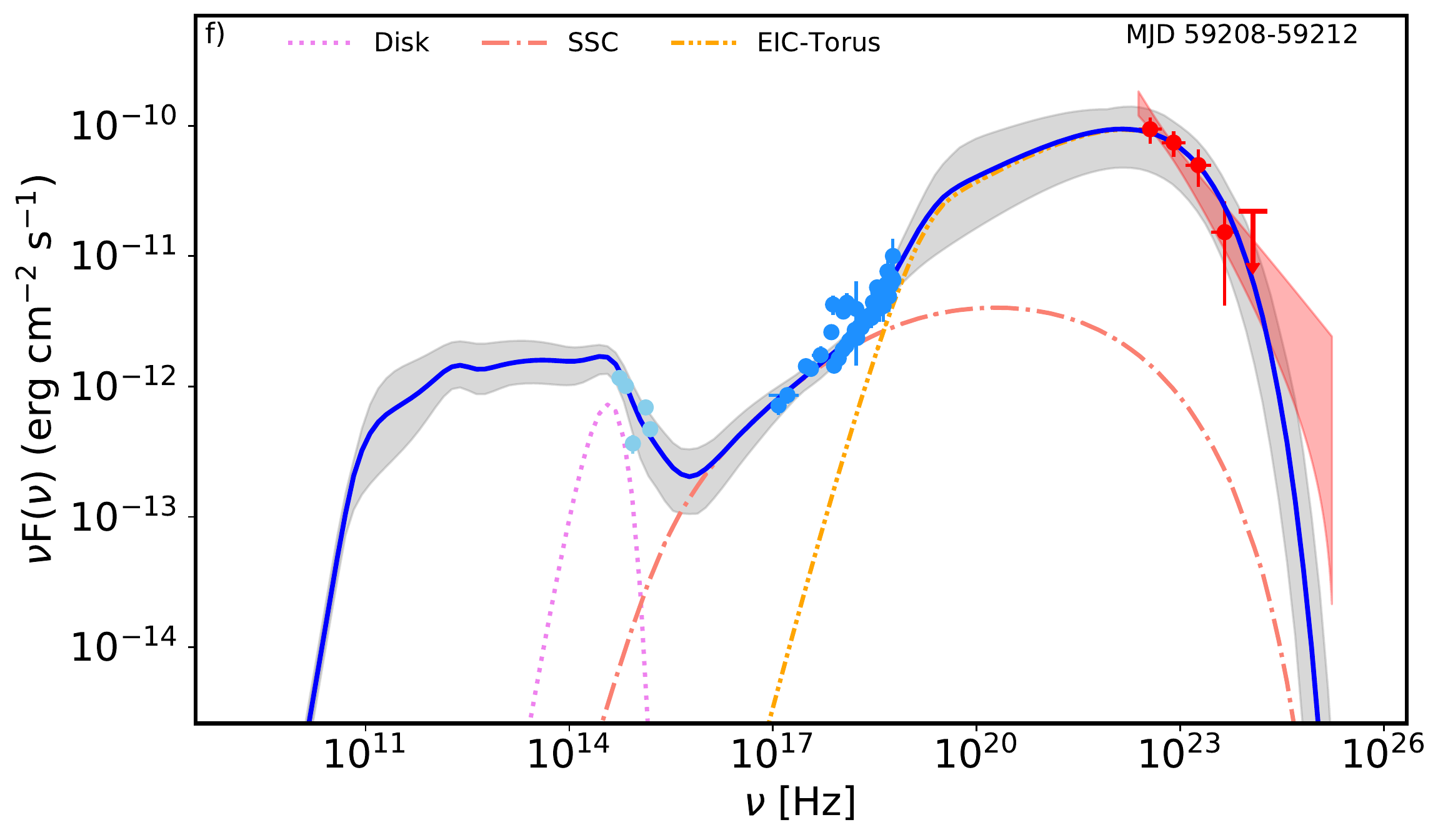}
    \caption{The broadband SEDs of \source\ in the quiescent (left panels) and flaring (right panels) states. Panels a and b correspond to SED modeling when the emission is entirely due to synchrotron/SSC radiation. The SED modeling when the emission region is within the BLR is shown in panels c and d, while in panels e and f it is outside the BLR. The blue solid curve shows the sum of all components and the gray shaded area is the uncertainty region from MCMC sampling of the parameters. The color code of the other components is given in the legends. In panels a and b, the observed and absorption-corrected Swift UVOT fluxes are shown with empty and filled light blue circles, respectively.}
    \label{SED}
\end{figure*}

Fig. \ref{SED} panels a and b show the results of the modeling when the entire emission is due to synchrotron/SSC emission from a compact region of the jet when the source is in a quiescent and flaring state, respectively. The corresponding model parameters are given in Table \ref{params}. In the quiescent state, the SED modeling shows that the spectral slope of the emitting particle distribution is $1.8\pm0.1$ and their distribution extends up to $(1.2\pm0.1)\times10^4$. The strength of the magnetic field is found to be $(9.3\pm0.8)\times10^{-3}$ G. The emission region size is $(2.0\pm0.1)\times10^{17}$ cm, which is consistent with the flux variability of $t_{\rm var}=(1+z)\times R/c\:\delta\approx18.7$ days. The Doppler boosting factor is $16.8\pm1.2$ which is not different from the values usually estimated for FSRQs \citep[e.g., see][]{2015MNRAS.448.1060G}. In this case, the synchrotron component decreases at $<10^{14}$ Hz and it does not take into account the observed optical/UV data which are interpreted as thermal emission from the accretion disk (see the next subsection).

In the flaring period (Fig. \ref{SED} panel b), the SED modeling shows that the emitting electrons have a harder spectrum with $p=1.6\pm0.03$ as compared with that in the quiescent state. The electrons are accelerated up to $\gamma_{\rm cut}=(1.1\pm0.1)\times10^4$ which is not significantly different from that in the quiescent state. In the flaring state, the magnetic field also increased, $B=(1.7\pm0.1)\times10^{-2}$ G, which is caused by the increase of the synchrotron flux. Also, the Doppler boosting factor increased from $16.8\pm1.2$ to $24.9\pm1.1$ in order to explain the slight shift of the HE peak towards higher energies; above 100 MeV the \gray\ spectrum in the flaring period has a photon index of $\Gamma_{\rm \gamma}=2.73\pm0.17$ as compared with that of $\Gamma_{\rm \gamma}=2.91\pm0.16$ is the quiescent state. The modeling shows that during the flare, the emission is produced from a smaller region with a radius of $(1.6\pm0.1)\times10^{17}$ cm corresponding to $t_{\rm var}\simeq10.0$ days, which indicates that the flaring emission is from a compact and faster moving region.  

\begin{table*}
    \centering
\caption{Parameters best describing the \source\ multiwavelength emission in the scenarios shown in Fig. \ref{SED}.}
\label{tab:example}
    \begin{tabularx}{\linewidth}{ *{7}{>{\centering\arraybackslash}X} }
    \multicolumn{1}{c}{} &
    \multicolumn{2}{c}{SSC} & \multicolumn{2}{c}{EIC-BLR} &
    \multicolumn{2}{c}{EIC Torus} \\
    \cmidrule(l){2-3}
    \cmidrule(l){4-5}
    \cmidrule(l){6-7}
     & quiescent       & flare    & quiescent     & flare  & quiescent     & flare  \\
    \midrule
    $p$ & $1.8\pm0.1$        & $1.6\pm0.03$       & $2.3\pm0.2$       & $2.2\pm0.1$ & $2.4\pm0.1$       & $2.4\pm0.1$       \\
    $\gamma_{\rm cut}/10^3$ & $11.8\pm0.8$        & $11.5\pm0.6$       & $0.3\pm0.1$       & $2.5\pm0.2$ & $1.3\pm0.1$       & $6.7\pm0.7$       \\
    $\gamma_{\rm min}$ & $9.5\pm1.0$        & $1.3\pm0.1$       & $26.1\pm3.8$       & $30.1\pm2.5$ & $65.7\pm3.4$       & $103.9\pm9.0$       \\
     $\delta$ & $16.8\pm1.2$        & $24.9\pm1.1$       & $13.4\pm1.3$       & $11.4\pm0.7$ & $15.3\pm0.7$       & $14.2\pm1.0$       \\
     $B [\rm G]$ & $(9.3\pm0.8)\times10^{-3}$        & $(1.7\pm0.1)\times10^{-2}$       & $3.5\pm0.4$       & $3.0\pm0.2$ & $0.2\pm0.01$       & $0.2\pm0.02$       \\
     $R [\rm cm]/10^{17}$ & $2.0\pm0.1$        & $1.6\pm0.1$       & $0.2\pm0.02$       & $0.1\pm0.01$ & $2.0\pm0.1$       & $1.4\pm0.1$       \\
     $L_{\rm e}[\rm erg\:s^{-1}]$ & $1.2\times10^{47}$        & $1.2\times10^{47}$       & $1.3\times10^{45}$       & $2.0\times10^{45}$ & $1.1\times10^{46}$       & $1.3\times10^{46}$       \\
     $L_{\rm B}[\rm erg\:s^{-1}]$ & $3.5\times10^{42}$        & $1.8\times10^{43}$       & $2.2\times10^{45}$       & $5.3\times10^{44}$ & $1.3\times10^{45}$       & $7.9\times10^{44}$       \\
     
    \bottomrule
  \end{tabularx}
  \label{params}
\end{table*}
\subsection{Emitting region within the BLR}
Fig. \ref{SED} panels c and d show the SED modeling assuming the jet dissipation occurred close to the central source. In this case, the density of disk-emitted and BLR-reflected photons in the jet frame (taking into account Doppler amplification) is comparable with or larger than that of synchrotron photons, so their inverse Compton scattering has a non-negligible contribution. The thermal emission from the accretion disk, modeled as a black body, is shown with a pink dashed line. In the quiescent state (Fig. \ref{SED} panel c), the low-energy component of the spectrum (up to $10^{16}$ Hz) can be reasonably well explained by combined synchrotron and black body components. The data in the X-ray band is mostly dominated by the EIC-disk component (blue dashed line in Fig. \ref{SED}) with SSC contributing in the soft X-ray band, whereas the emission in the \gray\ band is due to EIC-disk and EIC-BLR components (blue dashed and purple dot-dot-dashed lines in Fig. \ref{SED}, respectively). As compared with the synchrotron/SSC model, the distribution of the electrons is narrower with $p=2.3\pm0.2$ and $\gamma_{\rm cut}=(0.3\pm0.1)\times10^3$, because the average energy of the external photons is larger than that of the synchrotron one. The Doppler boosting factor is $\delta=13.4\pm1.3$ but the magnetic field is significantly larger, $B=3.5\pm0.4$ G. Since the flux in the HE band depends on the photon and particle density and the content of external photons is higher (inferred from the observed high Compton dominance, i.e., the ratio between the peak flux of inverse Compton and synchrotron components), the number of emitting electrons is reduced; to produce the synchrotron emission at the same level higher magnetic field is required. The emission is produced in a more compact region with a radius of $(0.2\pm0.02)\times10^{17}$ cm, smaller than $R_{\rm BLR}=9.3\times10^{17}$ cm.

During the flaring period, in the optical/UV band, the synchrotron emission from the jet dominates over the thermal emission from the accretion disk (Fig. \ref{SED} panel d). As the synchrotron  emission extends to higher frequencies, the SSC component makes a non-negligible contribution in the X-ray band (orange dot-dashed line in Fig. \ref{SED} panel d). The \gray\ emission is mostly due to the EIC-disk component (blue dashed line in Fig. \ref{SED} panel d) and EIC-BLR contributes at higher \gray\ energies (purple dot-dot-dashed line in Fig. \ref{SED} panel d). In this case, the electron distribution is nearly flat with $p=2.2\pm0.1$ and extends up to $(2.5\pm0.2)\times 10^3$. The increase of the energy up to which the electrons are effectively accelerated ($\gamma_{\rm cut}$) resulted in the shift of the synchrotron component to higher frequencies and domination over the disk thermal emission. The Doppler boosting is $\delta=11.4\pm0.7$, the magnetic field is $B=3.0\pm0.2$ G and the emission region radius is $(0.1\pm0.01)\times10^{17}$ cm. This is smaller than that estimated in the quiescent state and corresponds to $1.5$ days variability.
\subsection{Emitting region outside BLR}
Fig. \ref{SED} panels e and f show \source\ SED modeling assuming the emission region is beyond the BLR. In the quiescent state, the HE component is entirely dominated by EIC-torus (yellow dot-dot-dashed line in Fig. \ref{SED} panel e) and SSC contributing in the soft X-ray band (orange dot-dashed line in Fig. \ref{SED} panel e). In the flaring state, the peak of the SSC component is around $\sim10^{22}$ Hz (orange dot-dashed line in Fig. \ref{SED} panel f) making a non-negligible contribution to the X-ray band, but again, the HE \gray\ data is solely explained by the EIC-torus component (yellow dot-dot-dashed line in Fig. \ref{SED} panel f). The model parameters given in Table \ref{params} show that in the quiescent and flaring states the electron distribution has a similar power-law index $p\simeq2.4$, but in the flaring state the cut-off energy is larger, $\gamma_{\rm cut}=(6.7\pm0.7)\times10^3$ as compared to $\gamma_{\rm cut}=(1.3\pm0.1)\times10^3$. This is because \textit{i)} the synchrotron component should extend beyond the optical/UV band and \textit{ii)} during the flare the \gray\ spectrum is harder, shifting this component towards higher frequencies. Also, the modeling shows that the Doppler boosting and magnetic field do not substantially change, correspondingly being $\delta=15.3\pm0.7$ and $B=0.2\pm0.01$ G and $\delta=14.2\pm1.0$ and $B=0.2\pm0.02$ G for the flaring and quiescent states. However, again in the flaring state, the emission region has a slightly smaller radius $(1.4\pm0.1)\times10^{17}$ cm as compared with $(2.0\pm0.1)\times10^{17}$ cm.  
\subsection{Jet luminosity}
The parameters estimated during the modeling are used to compute the jet luminosity. The jet power carried by the electrons, calculated as $L_{e}=\pi c R_b^2 \Gamma^2 U_{e}$, and by magnetic field, calculated as $L_{B}=\pi c R_b^2 \Gamma^2 U_{B}$, are given in Table \ref{params}. In the case of synchrotron/SSC scenario (Fig. \ref{SED} panels a and b), the jet should be strongly particle dominated with a luminosity of the order of  $\simeq10^{47}\:{\rm erg\:s^{-1}}$ where the magnetic luminosity has a marginal contribution ($L_{\rm e}/L_{\rm B}\simeq3\times10^4$). This is natural, given the SED of \source\ in both quiescent and flaring periods shows strong Compton dominance. Relatively high luminosity is also estimated in the modeling when the emitting region is outside the BLR; ${\rm L_{\rm e}+L_{\rm B}=1.2\times10^{46}\:{\rm erg\:s^{-1}}}$ with $L_{\rm e}/L_{\rm B}=8.1$ and ${\rm L_{\rm e}+L_{\rm B}=1.4\times10^{46}\:{\rm erg\:s^{-1}}}$ with $L_{\rm e}/L_{\rm B}\simeq17$ for the quiescent and flaring states, respectively. When the emitting region is within the BLR, a lower jet luminosity is required, ${\rm L_{\rm e}+L_{\rm B}=6.3\times10^{45}\:{\rm erg\:s^{-1}}}$ and  ${\rm L_{\rm e}+L_{\rm B}=2.5\times10^{45}\:{\rm erg\:s^{-1}}}$ for the quiescent and flaring states, respectively, and the system is close to equipartition with $L_{\rm e}/L_{\rm B}=0.5$ and $L_{\rm e}/L_{\rm B}=3.7$ for the quiescent and flaring states, respectively.

Also, the total kinetic energy of the jet, defined as $L_{\rm kin}=L_{e}+L_{B}+L_{\rm p, cold}$, can be evaluated. Assuming a proton-to-electron comoving number density ratio of $N_{p}/N_{e}\simeq0.1$, in the most energy demanding model (synchrotron/SSC) $L_{\rm kin}=3.2\times10^{47}\:{\rm erg \: s^{-1}}$ and $L_{\rm kin}=4.4\times10^{47}\:{\rm erg \: s^{-1}}$ are estimated for the quiescent and flaring states, respectively. In the most optimistic scenario (EIC-BLR), $L_{\rm kin}=1.1\times10^{46}\:{\rm erg\:s^{-1}}$ and $L_{\rm kin}=6.0\times10^{45}\:{\rm erg\: s^{-1}}$ for the quiescent and flaring states, respectively. In this case, even if assuming $N_{p}/N_{e}\simeq1$, one would obtain $L_{\rm kin}=4.8\times10^{46}\:{\rm erg\: s^{-1}}$ and $L_{\rm kin}=3.8\times10^{46}\:{\rm erg\: s^{-1}}$ for the quiescent and flaring states, respectively. It is interesting that this luminosity is still lower than the disk luminosity estimated from the optical/UV data fitting. However, considering that the presence of the pairs can reduce the jet power \citep[e.g.,][]{2017MNRAS.465.3506P}, the estimated jet luminosity will be by several factors lower than the accretion disk luminosity.

In \citet{2010A&A...509A..69B}, the broadband SED of \source\ was modeled assuming the particles are injected into the emitting region, which is inside the BLR, and interact with the internal and external photon fields. The disk luminosity was estimated to be $1.7\times10^{47}\:{\rm erg\:s^{-1}}$ and $8.4\times10^{46}\:{\rm erg\:s^{-1}}$ by fitting the data observed in 2008 and 2006, respectively; the latter value is similar to the disk luminosity estimated in the current study. Their modeling showed that $L_{\rm e}$ is in the range of $(2.5-4.0)\times10^{46}\:{\rm erg\:s^{-1}}$ and $L_{\rm B}$ in  $(1.0-2.0)\times10^{45}\:{\rm erg\:s^{-1}}$. $L_{\rm e}$ is slightly larger than that estimated in the current study (see Table \ref{params}) which is related with different assumptions made in the modelings (e.g., emitting region radius, Doppler boosting factor, etc.).
\section{Summary}\label{summary}
In this work, we have performed a broadband study of the high redshift blazar \source. The main findings are summarized as follows:

\begin{itemize}
  \item \source\ is among the highest redshift blazars detected by \fermi. Its \gray\ emission, monitored since 2008, appeared relatively constant until 2017, then multiple powerful \gray\ flares were observed. Starting from MJD 59170, the source was in an enhanced \gray\ emission state when the \gray\ luminosity reached $6.14\times10^{49}\:{\rm erg\:s^{-1}}$. During the considered fourteen years, the \gray\ luminosity of the source exceeded $10^{49}\:{\rm erg\:s^{-1}}$ for 61.8 days in total. 
  \item The \gray\ photon index of the source varies as well. The mean of the \gray\ photon index during non flaring periods is $\simeq2.83$ which substantially hardens during the flares: the hardest index of $2.23\pm0.18$ was observed on MJD 59322. There is a moderate anti-correlation between the \gray\ photon index and luminosity.
  \item The source is very luminous in the X-ray band with a $0.3-10$ keV luminosity between $10^{47}-10^{48}\:{\rm erg\:s^{-1}}$ and with a hard spectrum ($\Gamma_{\rm X-ray}<1.38$). The available NuSTAR observations show that the hard X-ray spectrum extends up to 30 keV with $\Gamma_{\rm X-ray}=1.26$ with a luminosity between $(0.75-1.40)\times10^{48}\:{\rm erg\:s^{-1}}$.
  
  \item In order to understand the underlying physical processes at work in the jet of \source, the SEDs during the quiescent and flaring states were reproduced using a simple one-zone leptonic emission model considering different locations of the emission region. In the quiescent state, the combined synchrotron and thermal accretion disk components can explain the IR-optical-UV data, whereas X-ray to HE \gray\ data are due to inverse-Compton scattering of the disk and BLR-reflected photons. Instead, in the flaring state, the jet synchrotron emission dominates in the optical/UV band and the X-ray to HE \gray\ emission is due to combination of SSC, EIC-disk and EIC-BLR components. The modeling in the quiescent to flaring states showed that the flare was caused by the electron distribution changes, i.e., the electron power-law index hardened to $p=2.2\pm0.1$ and the cut-off energy was $\gamma_{\rm cut}=(2.5\pm0.2)\times10^3$.
  
  \item From the required jet energy point of view, the model with the emission region within the BLR is preferred. During the flaring event, the emitting region is nearly in equipartition with $L_{\rm e}/L_{\rm B}=3.7$ and the jet total luminosity is $L_{\rm tot}=3.8\times10^{46}\:{\rm erg\: s^{-1}}$ when assuming a proton-to-electron comoving number density ratio of $N_{p}/N_{e}\simeq1$. This luminosity is slightly lower than the accretion disk luminosity of $L_{\rm disc}=8.7\times10^{46}\:{\rm erg\:s^{-1}}$ estimated through fitting of UV/optical data.
  
\end{itemize}

Among the high red-shift blazars, \source\ is exceptional, having a reach multiwavelength data set (especially in the X-ray and \gray\ bands) which allows to investigate the processes taking place in the jet. Further multiwavelength monitoring of such distant and powerful sources will improve our understanding of the radiative processes at work in the relativistic jets in the early Universe.
\section*{Acknowledgements}
We acknowledge the use of data, analysis tools and services from the Open Universe platform, the ASI Space Science Data Center (SSDC), the Astrophysics Science Archive Research Center (HEASARC), the Fermi Science Tools, the Astrophysics Data System (ADS), and the National Extra-galactic Database (NED).

This work was supported by the Science Committee of the Republic of Armenia, in the frames of the research project No 21T-1C260.

This work used resources from the ASNET cloud.
\section*{Data availability}
All the data used in this paper is public and available from the Swift, Fermi and NuSTAR archives. The \fermi, Swift XRT/UVOT and NuSTAR data analyzed in this paper can be shared on a reasonable request to the corresponding author.



\bibliographystyle{mnras}
\bibliography{biblio} 




\appendix




\bsp	
\label{lastpage}
\end{document}